# First-Principles Simulation of Electron Mean-Free-Path Spectra and Thermoelectric Properties in Silicon

Bo Qiu[1], Zhiting Tian[1], Ajit Vallabhaneni[2], Bolin Liao[1], Jonathan M Mendoza[1], Oscar D. Restrepo[3], Xiulin Ruan[2], Gang Chen[1]*

1. Mechanical Engineering, Massachusetts Institute of Technology, Cambridge, MA, United States.

2. Mechanical Engineering, Purdue University, West Lafayette, IN, United States.

3. Materials Science and Engineering, Ohio State University, Columbus, OH, United States.

**Abstract**: The mean-free-paths (MFPs) of energy carriers are of critical importance to the nano-engineering of better thermoelectric materials. Despite significant progress in the first-principles-based understanding of the spectral distribution of phonon MFPs in recent years, the spectral distribution of electron MFPs remains unclear. In this work, we compute the energy dependent electron scatterings and MFPs in silicon from first-principles. The electrical conductivity accumulation with respect to electron MFPs is compared to that of the phonon thermal conductivity accumulation to illustrate the quantitative impact of nanostructuring on electron and phonon transport. By combining all electron and phonon transport properties from first-principles, we predict the thermoelectric properties of the bulk and nanostructured silicon, and find that silicon with 20 nm nanograins can result in more than five times enhancement in their thermoelectric figure of merit as the grain boundaries scatter phonons more significantly than that of electrons due to their disparate MFP distributions.

---

* Corresponding author: gchen2@mit.edu

Nanostructuring has proven to be an effective strategy to improve the figure of merit of thermoelectric materials [1–11]. The figure of merit is proportional to the electrical conductivity (σ), the square of the Seebeck coefficient (S) and inversely proportional to the thermal conductivity consisting of both phonon ($k_p$) and electron ($k_e$) contributions. The most effective nanostructuring approach so far has been reducing the phonon thermal conductivity while maintaining the electronic performance. For this strategy to be effective, the nanostructures should be smaller than the phonon mean free path (MFP) but larger than the electron MFP so that phonons are more strongly scattered than electrons. It is understood that both electron and phonon MFPs have a distribution over certain energy range. There has been good progress in predicting the spectral phonon MFPs for a range of bulk single crystals and alloys [12–24]. However, there has been no discussion on the spectral electron MFPs from first-principles. Surprisingly, this status exists even for silicon, one of the most important materials. Existing knowledge on electron scattering, relaxation time, and MFP, is mostly based on analytical models derived from Fermi's golden rule assuming ideal electron and phonon dispersions [25,26]. Past work on the phonon MFP distributions based on first-principles simulations, however, shows that such semi-empirical treatments on scattering lead to large error [13,21,23,27].

In this work, we compute the electron scattering rates and MFPs in silicon from first-principles and examine their dependence on energy, doping concentration, and their contributions to electronic conductivity and Seebeck coefficient. We demonstrate quantitatively the large disparity in the electron and phonon MFP distributions in silicon, and use the information obtained to predict that nanostructures with size of 20 nm can result in more than five times enhancement in ZT for silicon, consistent with past experimental results.

We consider n-doped silicon with carrier concentration between $10^{16}$ and $10^{19}$ cm$^{-3}$ in the temperature range 100 to 400 K. In this doping and temperature range, the dominant mechanisms for electron scatterings are electron-phonon and electron-impurity scatterings [28,29]. Both scattering rates are computed under the perturbation framework following Fermi's golden rule. The QUANTUM ESPRESSO package [30] is used to perform all density functional theory (DFT) calculations. In addition, the electron-phonon scattering is computed using the electron-phonon Wannier (EPW) package based on maximally localized Wannier functions [31,32], which allows accurate interpolation of electron-phonon couplings from coarse grids to arbitrarily dense grids [33]. The electron-impurity scattering is computed by explicitly accounting for the long range Coulomb tail for a screened ionized impurity, which was left out from previous first-principles-based works [28,34]. The Coulomb tail is described using classical model with inputs from first-principles. Inclusion of the long range Coulomb tail is very important to properly account for the effect of ionized dopants. The details of the computation of electron-phonon and impurity scattering rates, as well as transport coefficients are summarized in the Supplementary Information [35].

We first show the energy dependence of electron-phonon and electron ionized impurity scattering in Fig. 1(a). Fermi level corresponding to each carrier concentration is given in the corresponding legend. For silicon with n-type doping of ~$10^{19}$ cm$^{-3}$, the Fermi level is about right at the conduction band edge at 300 K. Hence, we plotted only results near the band edge. In semiconductors, the electron-phonon scattering is usually assumed to be of the form $E^{1/2}$ for acoustic phonon scattering [29] and Fig.1a shows that this approximation starts to break down for electron energy larger than 0.2 eV and smaller than 0.05 eV. The discrepancy for energies less than 0.05 eV is due to the omission of phonon energy in the energy conservation requirements in deriving the analytical expression [36], which leads to scattering rates that are proportional to the diminishing electron DOS near CBM. With phonon energies explicitly considered in this work, a finite scattering rate results instead. The discrepancy for energy larger than 0.2 eV relates to the deviation from parabolic band structure. The electron-phonon scattering also differs from analytical predictions for metals, which is of the form $E^{-3/2}$ [37]. The ionized impurity scattering for free electron scattering from a weakly screened ionized impurity is predicted to be of the form $E^{-3/2}$ while the first-principles results show that the exponent varies from -0.5 to -3.6, in comparison to -1.0 to -1.5 from empirical modeling [26], depending on the carrier concentration. The total electron scattering rate is obtained by summing both scattering channels following Matthiessen's rule $\tau_{el}^{-1} = \tau_{el-ph}^{-1} + \tau_{el-imp}^{-1}$, and is shown in Fig. 1(b). The total scattering rate follows the energy dependence of impurity and phonon scattering at low and high energies, respectively. At intermediate energy regimes, the energy dependence shows continuous transition from impurity to phonon scattering rates. The transition regime shifts to the high energies as impurity density increases. As temperature increases, more phonons are generated and the total scattering rate increases as well.

We compute thermoelectric transport properties based on the Boltzmann transport equation (BTE) and the obtained scattering rate. For example, the electrical conductivity can be expressed as [38]:

$$\sigma(T;\mu) = \frac{1}{V} \int \sum_{n\mathbf{k}} w_\mathbf{k} \, e^2 \, \Lambda_{n\mathbf{k}} v_{n\mathbf{k}} \delta(\varepsilon - \varepsilon_{n\mathbf{k}}) \left[ -\frac{\partial f_\mu(T;\varepsilon)}{\partial \varepsilon} \right] d\varepsilon \qquad (1)$$

where V is the crystal volume, w is the weighting factor, e is the elementary charge, ε is the electron energy, f is the Fermi-Dirac distribution, T is the temperature, and μ is the chemical potential. Λ is the energy dependent electron MFP obtained by multiplying the electron group velocity v and the energy dependent relaxation time τ:

$$\Lambda_{n\mathbf{k}}(\varepsilon) = v_{n\mathbf{k}}(\varepsilon) \cdot \tau_{n\mathbf{k}}(\varepsilon). \tag{2}$$

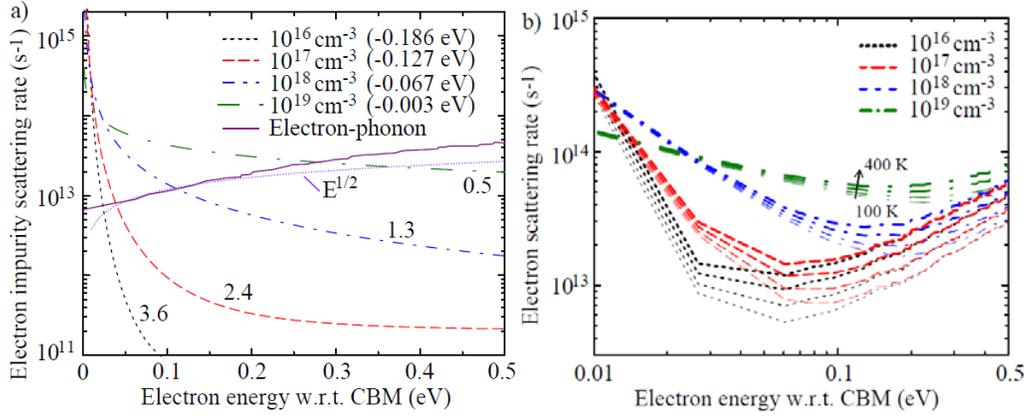

Figure 1. a) Electron-phonon and electron-impurity scattering rates with carrier concentration between $10^{16}$ and $10^{19}$ cm$^{-3}$ at 300 K. The chemical potential relative to the bandedge is shown inside the bracket. The numbers near the curves indicate the exponent $\alpha$ in $\tau^{-1} \propto E^{-\alpha}$. b) The combined electron-phonon and electron-impurity scattering rates.

We plot in Figs. 2 the dependence of the integrand in Eq. (1) for the electrical conductivity (Fig. 2a) and other thermoelectric properties. At higher temperatures, a wider span of electronic states contributes to transport due to the smearing of Fermi surface. For the electrical conductivity, the increase in carrier concentration dominates as carrier concentration increases, despite the increased electron scattering rate due to increased electron-impurity scattering. At higher impurity densities, the chemical potential shifts upwards. This leads to lowered average thermal energy for the carriers and, as a results, an overall decrease in the Seebeck coefficient, as seen in Fig. 2 (b). The double peak in Seebeck contribution at 100 K near 0.008 eV is due to numerical error from the band-crossing and it does not affect results, as seen in Fig. 2(d). From Fig. 2(c) and 2(d), it can be seen that the transport is mostly contributed from electronics with energies within 0.2 eV from CBM. The dominant contribution shifts to higher energies as temperature and carrier concentration increase.

The energy dependent electron MFPs are shown in Fig. 3. In general, the electron MFPs depend strongly on both temperature and carrier concentration. The MFPs monotonically decrease with increasing temperatures mainly due to increased electron-phonon scattering rates, which become less important at higher impurity densities. The increase in carrier concentration leads to monotonic decrease in the MFPs due to the domination of electron-impurity scattering. As seen in Fig. 3, at low temperature and low carrier concentration, the electron MFPs can go up to about 100 nm in silicon at 100 K. The MFPs reduce to about 20 nm with carrier concentration of $10^{19}$ cm$^{-3}$ at 400 K. The MFPs vanish as electron energy approaches CBM. This is because the band structure is parabolic near CBM in silicon that the band velocity approaches zero as electron energy approaches CBM, leading to vanishing MFPs. It should also be pointed out that, unlike the distribution of phonon MFP, which spans several orders of magnitude in silicon and tends to diverge at low phonon frequencies [13], the electron MFP in silicon shows much narrower span and does not diverge throughout the energy range relevant to the electron transport.

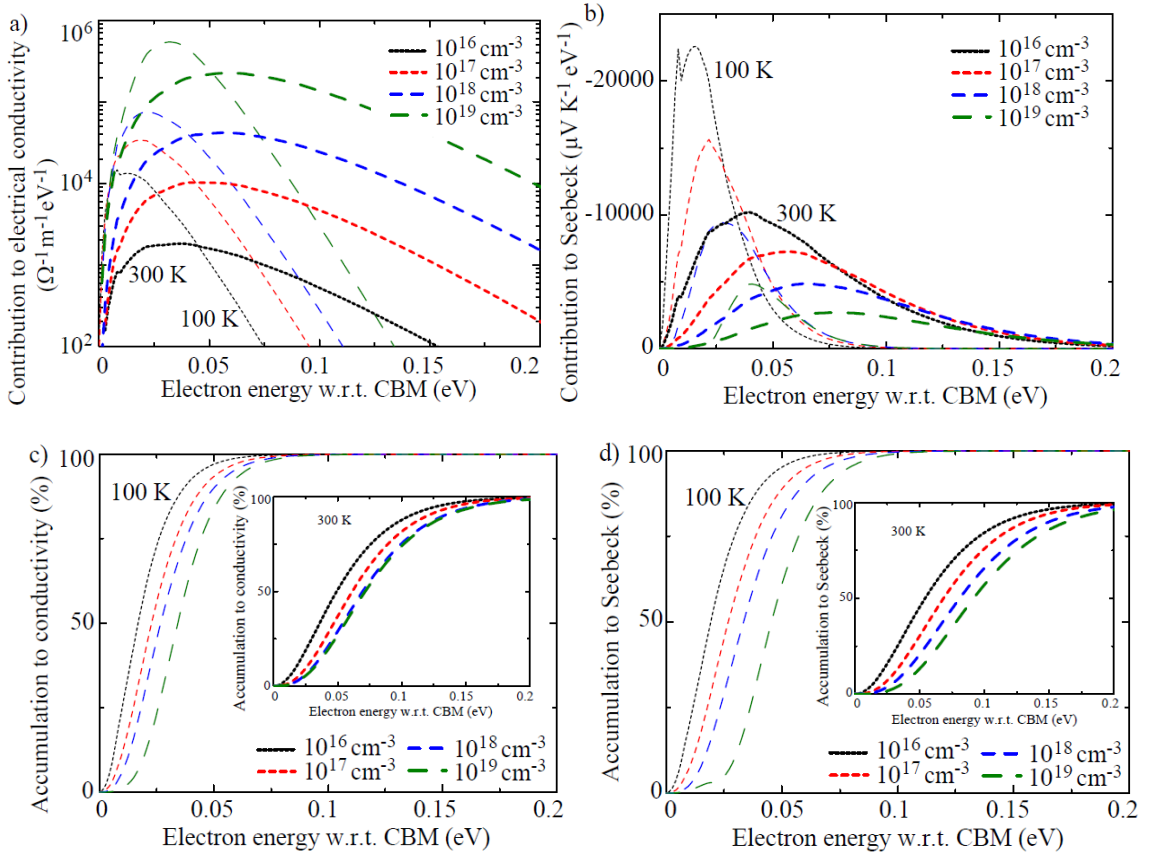

Figure 2. a) Per-energy interval contribution to electrical conductivity. b) Per-energy interval contribution to the numerator of the Seebeck coefficient divided by the corresponding electrical conductivity. c) Percentage accumulation to electrical conductivity. d) Percentage accumulation to Seebeck coefficient.

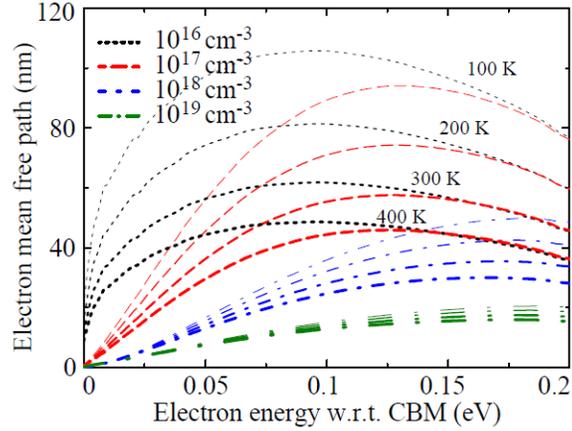

Figure 3. Electron MFP as a function of impurity density and temperature. Thin to thick curves indicate temperatures from 100 to 400 K.

To understand how energy carriers are affected by nanostructuring, it is useful to make an accumulation plot of electrical/thermal conductivity with respect to electron/phonon MFPs. The accumulation of phonon thermal conductivity with MFP has been extensively studied and has been documented elsewhere [13,39]. The contribution to electrical conductivity from electrons with MFPs up to $\Lambda$ can be obtained according to Eq. (1) by summing over contributions from all electrons with MFPs less than that of $\Lambda$:

$$\sigma(\Lambda) = \frac{e^2}{N_{\mathbf{k}}V} \sum_{n\mathbf{k}}^{\Lambda_{n\mathbf{k}}<\Lambda} \left[ -\frac{\partial f_\mu(T;\varepsilon)}{\partial \varepsilon} \right]_{\varepsilon_{n\mathbf{k}}} v_{n\mathbf{k}} \Lambda_{n\mathbf{k}} . \quad (3)$$

Here $N_k$ is the number of k points. The electron MFP contribution is weighted strongly by the derivative of the Fermi-Dirac distribution. As a result, only electrons with energy falling into the Fermi window contribute mostly to the electrical conductivity.

The accumulation plots for both electrons and phonons in silicon are compared in Fig. 4. The electron MFPs show much narrower span in comparison to phonon MFPs. At $10^{19}$ cm$^{-3}$ doping, majority of electrical conductivity contribution comes from electrons with MFPs less than 10 nm.

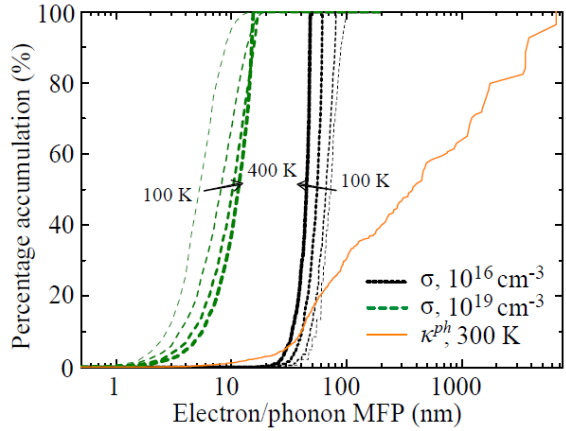

Figure 4. Electrical and lattice thermal conductivity accumulation with respect to MFPs. Thin to thick curves indicate temperatures from 100 to 400 K.

The trends of temperature dependence of the MFPs are opposite for carrier concentrations of $10^{16}$ and $10^{19}$ cm$^{-3}$. At $10^{16}$ cm$^{-3}$, the MFP shortens at elevated temperatures due to increased scattering. At $10^{19}$ cm$^{-3}$, the main contribution to electrical conductivity shifts to longer MFP at higher temperatures. This is due to the broadened Fermi window that includes more contribution from electrons with longer MFPs. Note that the phonon accumulation shown here only includes phonon-phonon scattering effects while electron-phonon [40] and phonon-impurity [20] scatterings are not considered in this work. The inclusion of electron-phonon scattering can lead to ~7% reduction in overall lattice thermal conductivity at electron concentration of $10^{19}$ cm$^{-3}$ [40].

Good thermoelectric materials require low thermal conductivity and good electrical conductivity. According to Fig. 4, nanostructures with size about 20 nm can be chosen to maximally scatter phonons while retaining the majority of electron transport at carrier concentrations of $10^{19}$ cm$^{-3}$. To evaluate the effect of nanostructuring in Si, we re-compute the thermoelectric transport coefficients by adding the boundary scattering rates ($\tau_b^{-1} = v/D$) to the bulk scattering rates following Matthiessen's rule for both electrons and phonons. In Fig.5, we show simulated thermoelectric figure of merit for both bulk Si and nanocrystalline Si, and compare with available experimental data. For nanocrystalline Si, we take d=20 nm to be close to experimentally reported average grain size. The figure shows that simulation results are in reasonable agreement with experiments for both bulk Si crystals and bulk nanocomposites [41]. The predicted ZT of the nanostructured bulk Si is five times higher than that of bulk single crystals. We point out that experimental data are available for carrier concentration at $10^{20}$ cm$^{-3}$ and our simulation can only reach $10^{19}$ cm$^{-3}$ [35]. At the carrier concentration of $10^{20}$ cm$^{-3}$, less electrical conductivity reduction can be expected due to shorter bulk electron MFP and weaker boundary scattering, which may translate into higher ZT.

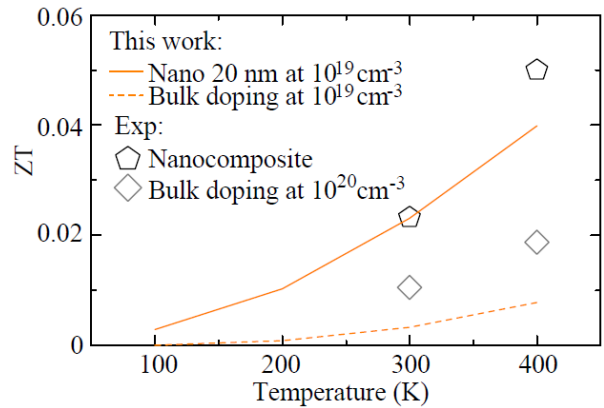

Figure 5. Predicted enhancement in thermoelectric figure of merit ZT due to nanostructuring. Experiments are from Ref. [41].

In summary, we have computed electron scatterings and MFPs in silicon from first-principles, and found that the energy dependence of ionized impurity scattering rate differs significantly from existing analytical expressions. Our simulations show large disparity of electron and phonon MFPs distributions that favor nanoengineering in silicon to improve thermoelectric figure of merit ZT. We predicted that nanostructures with sizes of 20 nm can result in more than five times enhancement in ZT for silicon doped at $10^{19}$ cm$^{-3}$, owing to the strong scattering of phonons and less-affected electron transport, and the results are in reasonable agreement with past experiment. Our work shows the potential of using first-principles tools in engineering nanostructures for thermoelectric energy conversion.

We would like to thank Dr. David Singh, Dr. Feliciano Giustino and Dr. Vincenzo Lordi for insightful discussions. This work was supported as part of the Solid-State Solar-Thermal Energy Conversion Center (S3TEC), an Energy Frontier Research Center funded by the US Department of Energy, Office of Science, Office of Basic Energy Sciences (Grant No. DE-SC0001299). O. D. R. acknowledges support from the Center for Emergent Materials at The Ohio State University, an NSF MRSEC (Grant No. DMR-0820414).

# Supplemental Information for First-Principles Simulation of Electron Mean-Free-Path Spectra and Thermoelectric Properties in Silicon

Bo Qiu[1], Zhiting Tian[1], Ajit Vallabhaneni[2], Bolin Liao[1], Jonathan M Mendoza[1], Oscar D. Restrepo[3], Xiulin Ruan[2], Gang Chen[1*]

1. Mechanical Engineering, Massachusetts Institute of Technology, Cambridge, MA, United States.

2. Mechanical Engineering, Purdue University, West Lafayette, IN, United States.

3. Materials Science and Engineering, Ohio State University, Columbus, OH, United States.

## I. Methodology

The efficiency of a thermoelectric device is determined by the figure of merit

$$ZT = \frac{S^2 \sigma T}{\kappa^{el} + \kappa^{ph}} \quad . \tag{1}$$

In order to predict $ZT$, the electron thermoelectric coefficients Seebeck coefficient $S$, electrical conductivity $\sigma$, and electronic thermal conductivity $\kappa^{el}$ as well as the lattice thermal conductivity $\kappa^{ph}$ need to be evaluated. As detailed in the following sections, BTE will be used to evaluate these transport coefficients. The electronic band structure and phonon dispersion relations will first be computed using density functional theory (DFT) and density functional perturbation theory (DFPT), respectively. Such computations do not present significant challenges for many single-element materials and common non-transition-metal compounds. They can be done mostly routinely thanks to the advancement in a variety of computational packages capable of DFT, DFPT, and frozen-phonon calculations, such as QUANTUM ESPRESSO, WIEN2K, ABINIT, VASP, and so on. On the other hand, the computation of carrier scattering is far more challenging and is in fact the main hurdle for a first-principles prediction of $ZT$. Here we provide a first-principles framework to evaluate carrier scattering rates based on the perturbation theory and Fermi's golden rule.

### A. Transport coefficients from BTE

In the presence of an electrical field or temperature gradient, electrical and thermal currents will be generated in materials. The relationship between electrical field and temperature gradient and the corresponding electrical current $J$ and thermal current $J^Q$ are [1,2]

$$\begin{aligned} J_\alpha &= \sigma_{\alpha\beta} E_\beta - N_{\alpha\beta} \nabla_\beta T \\ J_\alpha^Q &= T N_{\alpha\beta} E_\beta - k_{\alpha\beta}^0 \nabla_\beta T \end{aligned} \quad . \tag{2}$$

---

Here $E$ is the electromotive force combining electrostatic field and chemical potential gradient, and $T$ is the temperature, and the subscript $\alpha = x, y, z$ represents Cartesian coordinate components. We use Einstein notation where summation over repeated indices are implicit that $\sigma_{\alpha\beta} E_\beta \equiv \sum_\beta \sigma_{\alpha\beta} E_\beta$. The Seebeck coefficient $S$ is defined as the resultant voltage gradient produced by a temperature gradient at zero electric current:

$$S_{\alpha\beta} = (\sigma^{-1})_{\alpha\gamma} N_{\gamma\beta}. \tag{3}$$

The electronic thermal conductivity is defined at zero electric current:

$$k_{\alpha\beta}^{el} = k_{\alpha\beta}^0 - T N_{\alpha\gamma} (\sigma^{-1})_{\gamma\delta} N_{\delta\beta}. \tag{4}$$

The electronic thermal conductivity is often related to the electrical conductivity through the Wiedemann-Franz (W-F) law

$$\kappa_{\alpha\beta}^{el} = L_0 \sigma_{\alpha\beta} T. \tag{5}$$

where $L_0$ is the Lorenz number and is a constant for metal $L_0 = 2.44 \times 10^{-8}$ WΩK$^{-2}$, for semiconductors, it depends on carrier concentration and will be discussed later.

To evaluate the above electronic transport coefficients, the key is to model the electrical conductivity tensor. For general electron transport, the electrical current of carriers is defined as

$$J_\alpha = e \sum_{n\mathbf{k}} f_{n\mathbf{k}} v_{n\mathbf{k}}^\alpha, \tag{6}$$

where $f_{n\mathbf{k}}$ is the population of electronic state $n\mathbf{k}$ under electrochemical and temperature gradients, the group velocity is

$$v_{n\mathbf{k}}^\alpha = \frac{1}{\hbar} \frac{\partial \varepsilon_{n\mathbf{k}}}{\partial k_\alpha}, \tag{7}$$

where $\varepsilon_{n\mathbf{k}}$ is the energy of an electronic state that is taken as the eigenvalue obtained by solving the Kohn-Sham (KS) equation self-consistently [3]. $k_\alpha$ is the $\alpha$-th component of wavevector $\mathbf{k}$ The BTE is formulated [1] to solve for the carrier distribution function $f_{n\mathbf{k}}$, which is the only other unknown to evaluate the electrical current in Eq. (6)

$$\frac{\partial f_{n\mathbf{k}}}{\partial t} + v_{n\mathbf{k}}^\alpha \frac{\partial f_{n\mathbf{k}}}{\partial r_\alpha} + \frac{e}{\hbar} \left( E_\alpha + \frac{1}{c} (v_{n\mathbf{k}} \times \mathbf{H})_\alpha \right) \frac{\partial f_{n\mathbf{k}}}{\partial k_\alpha} = \left. \frac{df_{n\mathbf{k}}}{dt} \right|_{scat}. \tag{8}$$

In the absence of magnetic field and temperature gradient, by using relaxation time approximation (RTA) for the scattering term

$$\left.\frac{df_{n\mathbf{k}}}{dt}\right|_{scat} \approx -\frac{f_{n\mathbf{k}} - f_0}{\tau_{n\mathbf{k}}}, \tag{9}$$

the BTE can then be linearized as

$$f_{n\mathbf{k}} = f_0 + e\left(-\frac{\partial f_0}{\partial \varepsilon}\right)\tau_{n\mathbf{k}} v_{n\mathbf{k}}^{\alpha} E_{\alpha} . \tag{10}$$

Here $\tau_{n\mathbf{k}}$ is the relaxation time that depends on both the band index n and wavevector **k** and the local equilibrium Fermi-Dirac distribution for electrons that depends on local electrochemical potential and temperature. By inserting Eq. (10) into Eq. (6), the electrical conductivity is obtained as

$$\sigma_{\alpha\beta} = e^2 \sum_{n\mathbf{k}} \left(-\frac{\partial f_0}{\partial \varepsilon}\right) \tau_{n\mathbf{k}} v_{n\mathbf{k}}^{\alpha} v_{n\mathbf{k}}^{\beta} . \tag{11}$$

If we define the transport kernel as

$$\Xi_{n\mathbf{k}}^{\alpha\beta} = e^2 \tau_{n\mathbf{k}} v_{n\mathbf{k}}^{\alpha} v_{n\mathbf{k}}^{\beta} , \tag{12}$$

the energy projected transport kernel can be obtained as

$$\Xi_{\alpha\beta}(\varepsilon) = \sum_{n\mathbf{k}} w_{\mathbf{k}} \Xi_{n\mathbf{k}}^{\alpha\beta} \delta(\varepsilon - \varepsilon_{n\mathbf{k}}) , \tag{13}$$

where $w_{\mathbf{k}} = 1/N_{\mathbf{k}}$ is the weighting factor and $N_{\mathbf{k}}$ is the number of **k** points due to discretization. $\delta(\varepsilon - \varepsilon_{n\mathbf{k}})$ is the Dirac delta function. Following the procedure similar to deriving the electrical conductivity, all electron thermoelectric transport coefficients can be obtained as

$$\begin{aligned}
\sigma_{\alpha\beta}(T;\mu) &= \frac{1}{V}\int \Xi_{\alpha\beta}(\varepsilon)\left[-\frac{\partial f_{\mu}(T;\varepsilon)}{\partial \varepsilon}\right]d\varepsilon \\
N_{\alpha\beta}(T;\mu) &= \frac{1}{eTV}\int (\varepsilon-\mu)\Xi_{\alpha\beta}(\varepsilon)\left[-\frac{\partial f_{\mu}(T;\varepsilon)}{\partial \varepsilon}\right]d\varepsilon \\
S_{\alpha\beta}(T;\mu) &= \frac{(\sigma^{-1})_{\alpha\gamma}(T;\mu)}{eTV}\int (\varepsilon-\mu)\Xi_{\gamma\beta}(\varepsilon)\left[-\frac{\partial f_{\mu}(T;\varepsilon)}{\partial \varepsilon}\right]d\varepsilon \\
\kappa_{\alpha\beta}^{0}(T;\mu) &= \frac{1}{e^2 TV}\int (\varepsilon-\mu)^2 \Xi_{\alpha\beta}(\varepsilon)\left[-\frac{\partial f_{\mu}(T;\varepsilon)}{\partial \varepsilon}\right]d\varepsilon
\end{aligned} \tag{14}$$

Here $\mu$ is the chemical potential and $V$ is the crystal volume. The electron mobility $\mu_n$ can be obtained as

$$\mu_n(T;\mu) = \frac{\sigma_{\alpha\beta}(T;\mu)}{n(T;\mu)e}. \tag{15}$$

As seen, to evaluate all electron thermoelectric coefficients, the key is to evaluate the transport kernel in Eq. (12) from first-principles. Besides the band energy and group velocities that can be readily obtained from standard DFT calculations, the major challenge lies in the evaluation of the electron relaxation times from first-principles. The first-principles evaluation of electron-phonon and electron-impurity scatterings will be detailed in the following sections.

Although first-principles based simulation of phonon thermal conductivity is well established [4–6], for completeness in describing thermoelectric properties and ZT calculation, we will give below a short summary. The BTE for an individual phonon mode can be written as [1]

$$v_{\mathbf{q}\nu}^\alpha \nabla_\alpha n_{\mathbf{q}\nu} = \left.\frac{dn_{\mathbf{q}\nu}}{dt}\right|_{scat}. \tag{16}$$

Again, in the case of small deviation of the phonon population $n_{\mathbf{q}\nu}$ from equilibrium Bose-Einstein distribution $n_0$, we can approximate the scattering term with RTA:

$$\left.\frac{dn_{\mathbf{q}\nu}}{dt}\right|_{scat} \approx -\frac{n_{\mathbf{q}\nu} - n_0}{\tau_{\mathbf{q}\nu}}. \tag{17}$$

By substituting Eq. (17) into Eq. (16) and applying to the expression of general heat flux

$$q_\alpha = \sum_{\mathbf{q}\nu} \frac{\hbar\omega}{V} v_{\mathbf{q}\nu}^\alpha (n_{\mathbf{q}\nu} - n_0) \tag{18}$$

then applying Fourier's law, the lattice thermal conductivity can be expressed as

$$\kappa_\alpha^{ph} = \frac{-q_\alpha}{\partial T/\partial x} = \sum_{\mathbf{q}\nu} c_{\mathbf{q}\nu}^{ph} \left(v_{\mathbf{q}\nu}^\alpha\right)^2 \tau_{\mathbf{q}\nu}. \tag{19}$$

Here

$$c_{\mathbf{q}\nu}^{ph} = \frac{1}{V}\frac{\partial\left(n_{\mathbf{q}\nu}\hbar\omega_{\mathbf{q}\nu}\right)}{\partial T} = \frac{k_B \vartheta}{V}\frac{\exp(\vartheta)}{\left[\exp(\vartheta)-1\right]^2}. \tag{20}$$

is the specific heat capacity per phonon mode and $\vartheta \equiv \hbar\omega_{\mathbf{q}\nu}/k_B T$.

Again, similar to the case of electronic transport coefficients, to evaluate the lattice thermal conductivity from first-principles, the key challenge is to evaluate the phonon relaxation time.

As mentioned above, we use the RTA to formulate transport coefficients for electrons and phonons. However, it was also pointed out that the zeroth-order solution to BTE as RTA may lead to inaccuracy when the system is far from equilibrium and a fully iterative approach should be used [7], such as in the case of high electric field or momentum gradient. At low or intermediate temperatures for non-degenerate semiconductors, the RTA has been shown to well predict electron transport for a wide range of materials [8,9]. For phonon transport, it was pointed out by Broido and colleagues that for materials where normal processes (N-process) dominate in materials such as diamond, graphene and CNT, the RTA can significantly underestimate the lattice thermal conductivity [5,10,11]. This is attributed to the fact that RTA treats all scattering paths as equally resistive while the N-process only indirectly contributes to resistivity by redistributing the carriers for Umklapp scattering (U-process). In these instances, a fully iterative solution is needed. Nonetheless, the use of RTA is accurate enough to reproduce the phonon transport in many common materials such as silicon [4].

### B. Electron-phonon scattering

As detailed in Appendix A, according to Fermi's golden rule and under the RTA, the electron-phonon scattering rate, or the inverse of electron relaxation time limited by phonon scattering, can be obtained as

$$\frac{1}{\tau_{n\mathbf{k}}} = \frac{2\pi}{\hbar} \sum_{\mathbf{q}\nu} w_{\mathbf{q}} |g_{mn}^{\nu}(\mathbf{k},\mathbf{q})|^2 \left[ \begin{array}{l} \left(n(\omega_{\mathbf{q}\nu}) + f(\varepsilon_{m\mathbf{k}+\mathbf{q}})\right) \delta(\varepsilon_{n\mathbf{k}} - \varepsilon_{m\mathbf{k}+\mathbf{q}} + \hbar\omega_{\mathbf{q}\nu}) \\ + \left(n(\omega_{\mathbf{q}\nu}) + 1 - f(\varepsilon_{m\mathbf{k}+\mathbf{q}})\right) \delta(\varepsilon_{n\mathbf{k}} - \varepsilon_{m\mathbf{k}+\mathbf{q}} - \hbar\omega_{\mathbf{q}\nu}) \end{array} \right]. \quad (21)$$

Here the first and second term represents the transition out of state $n\mathbf{k}$ by absorption and emission of a phonon, respectively. The scattering rate is related to the imaginary part of electron self-energy within the Migdal approximations as [12]:

$$\frac{1}{\tau_{n\mathbf{k}}} = \frac{2}{\hbar} \text{Im}(\Sigma_{n\mathbf{k}}), \quad (22)$$

where

$$\text{Im}(\Sigma_{n\mathbf{k}}) = \text{Im}\left( \sum_{\mathbf{q}\nu} w_{\mathbf{q}} |g_{mn}^{\nu}(\mathbf{k},\mathbf{q})|^2 \left[ \frac{n(\omega_{\mathbf{q}\nu}) + f(\varepsilon_{m\mathbf{k}+\mathbf{q}})}{\varepsilon_{n\mathbf{k}} - \varepsilon_{m\mathbf{k}+\mathbf{q}} + \hbar\omega_{\mathbf{q}\nu} - i\eta} + \frac{n(\omega_{\mathbf{q}\nu}) + 1 - f(\varepsilon_{m\mathbf{k}+\mathbf{q}})}{\varepsilon_{n\mathbf{k}} - \varepsilon_{m\mathbf{k}+\mathbf{q}} - \hbar\omega_{\mathbf{q}\nu} - i\eta} \right] \right). \quad (23)$$

Here the summation goes over all possible phonon modes $\mathbf{q}\nu$ that can scatter electrons while satisfying conservation of energy and crystal momentum. $w_{\mathbf{q}} = 1/N_{\mathbf{q}}$ is the weighting factor used to normalize the contribution to scattering rate from a small reciprocal space volume adjacent to $\mathbf{q}$ due to the choice of discretized phonon q-grid. $f$ and $n$ are the Fermi-Dirac and Bose-Einstein distribution for electrons and phonons, respectively. $\varepsilon$ and $\hbar\omega$ are the energies of electron and phonon states, respectively. $\eta$ is the Gaussian broadening parameter. Within the DFPT framework, the electron-phonon matrix element [13]

$$g_{mn}^{\nu}(\mathbf{k},\mathbf{q}) = \sqrt{\frac{\hbar}{2\omega_{\mathbf{q}\nu}}} \langle \psi_{m\mathbf{k}+\mathbf{q}} | \delta U_{\mathbf{q}\nu} | \psi_{n\mathbf{k}} \rangle \tag{24}$$

describes the electron-phonon coupling strength in the event of the scattering of an electron from a Bloch state $n\mathbf{k}$ to another state $m\mathbf{k}+\mathbf{q}$ by a phonon mode $\mathbf{q}\nu$ with frequency $\omega_{\mathbf{q}\nu}$. Here $\psi$ is the wavefunction of an electronic ground eigenstate and

$$\delta U_{\mathbf{q}\nu} = \sum_{\mathbf{R}} \sum_{s} \mathbf{e}_{\mathbf{q}\nu}^{s} \cdot \frac{\partial U}{\partial \mathbf{u}_{\mathbf{R}s}} \frac{e^{i\mathbf{q}\cdot\mathbf{R}}}{\sqrt{N}} \tag{25}$$

is the first order variation of the Kohn-Sham (KS) potential that is composed of first-order derivative of the self-consistent KS potential $U$ with respect to the atomic displacements of s-th atom in lattice position $\mathbf{R}$. Here

$$\mathbf{e}_{\mathbf{q}\nu}^{s} = \frac{\mathbf{u}_{\mathbf{q}\nu}^{s}}{\sqrt{M_s}} \tag{26}$$

are the is the phonon displacement vectors that are proportional to the phonon eigenvectors $\mathbf{u}_{\mathbf{q}\nu}^{s}$, which are the eigenvectors of the dynamical matrix. The deformation potential is proportional to the modulus of the electron-phonon matrix element in Eq. (24). The commonly referred effective deformation potential can be obtained as the average of the k and q-dependent deformation potentials [14].

To evaluate the electron-phonon scattering rate, the knowledge of electron and phonon band structures are required so all possible scattering events can be screened by requiring momentum and energy conservation:

$$\mathbf{k}' = \mathbf{k} \pm \mathbf{q} + \mathbf{G} \quad , \quad \varepsilon_{m\mathbf{k}'} = \varepsilon_{n\mathbf{k}} \pm \hbar\omega_{\mathbf{q}\nu} \; . \tag{27}$$

Here $\mathbf{G}$ is a reciprocal lattice vector.

We use the QUANTUM ESPRESSO package [15] to carry out all DFT and DFPT calculations. Using the pseudopotential approach, the KS equation is solved iteratively until specified convergence criteria are reached to obtain the self-consistent electron eigenvalues and wavefunctions. The phonon frequencies and eigenvectors are computed using the linear-response theory under the DFPT framework. The second order change in the total energy, which is directly related to the dynamical matrix, is obtained by using the variational principle (2n+1 theorem) and up to first order wavefunction or the first order change in electron density. The linear response of the electronic density to atomic displacement is obtained by solving a set of self-consistent equations for the perturbed system analogous to the KS equations [16].

While the calculations of ground state electron properties and harmonic vibrational properties are usually computationally affordable, the main challenges lie in the evaluation of the electron-phonon matrix elements. To evaluate the matrix elements, only the ground state self-consistent wavefunctions and the first order derivative of KS potential are needed. However, the computation cost is proportional to the number of **k** points of an electron grid multiplied by the number of **q** points of a phonon grid. Since dense electron and phonon grids are usually required to achieve the convergence of scattering rates, brute force computation of coupling among the full electron and phonon grid is presently prohibitive.

To achieve numerical results on a dense enough combined grid, interpolation techniques are usually necessary. Linear interpolation has been used to obtain the electron-phonon matrix elements on a denser electron-phonon grid [9,17], requiring affordable cost for computation of the matrix elements on a much coarser grid. However, for generality, an efficient approach without sacrificing accuracy in the interpolation of matrix elements is needed, especially for thermoelectric materials that are likely to have complicated electron-phonon couplings. Therefore, we choose to use the Wannier interpolation approach recently developed in Louie and Cohen's groups [18,19]. The essential idea of this approach is to utilize the localization of both electronic and lattice MLWF to achieve accurate interpolation of arbitrarily dense electron-phonon matrix elements. The electron and phonon eigenvalues, eigenvectors, and the electron-phonon matrix elements are first computed on a coarse grid in the Bloch space. Then they are transformed into Wannier space through MLWF. In the Wannier representation, if the two electron Wannier functions centering at different unit cells are sufficiently apart, the electron-phonon matrix elements in Wannier representation vanishes. Since the electron Hamiltonian, phonon dynamical matrix, and electron-phonon matrices decay spatially by construction, one needs only a small number of elements in the Wannier representation to be complete. In principle, these elements are enough to allow the Fourier transformation back to the Bloch space to obtain electron bandstructure, phonon dispersions, and electron-phonon matrices on arbitrarily dense grids with high accuracy.

To further illustrate the idea of Wannier interpolation, we briefly describe the above-mentioned procedure for electrons. The electron Hamilton in Bloch space is

$$H^{el}_{mn,\mathbf{k}} = \langle m\mathbf{k} | \hat{H}^{el} | n\mathbf{k} \rangle . \tag{28}$$

The corresponding Hamiltonian in the Wannier representation is [19]:

$$H^{el}_{mn;\mathbf{R}_e,\mathbf{R}'_e} = \langle \mathbf{R}_e | \hat{H}^{el} | \mathbf{R}'_e \rangle = \sum_{\mathbf{k}} e^{-i\mathbf{k}\cdot(\mathbf{R}'_e - \mathbf{R}_e)} M^{\dagger}_{\mathbf{k}} H^{el}_{mn,\mathbf{k}} M_{\mathbf{k}} , \tag{29}$$

where $M_{\mathbf{k}}$ is the unitary function constructed using electron eigenvalues and eigenvectors to enable transformation into MLWF. $H^{el}_{\mathbf{R}_e,\mathbf{R}'_e}$ decays within $|\mathbf{R}_e - \mathbf{R}'_e|$. Therefore one can assume and confirm numerically that for $\mathbf{R}_e$ outside a Wigner-Seitz volume, $H^{el}_{\mathbf{0}_e,\mathbf{R}'_e}$ vanishes. Then in the transform from Wannier representation back to the Block space

$$H^{el}_{mn,\mathbf{k'}} = M_{\mathbf{k'}} \left( \frac{1}{N_e} \sum_{\mathbf{R}_e} e^{i\mathbf{k'}\cdot\mathbf{R}_e} H^{el}_{mn;\mathbf{0}_e,\mathbf{R}_e} \right) M^{\dagger}_{\mathbf{k'}} . \qquad (30)$$

To obtain $H^{el}_{mn,\mathbf{k'}}$ for arbitrary wavevector $\mathbf{k'}$, the summation over only a finite number of elements $H^{el}_{\mathbf{0}_e,\mathbf{R}_e}$ in the Wannier representation within a small spatial cutoff is sufficient.

In this work, the electron-phonon couplings, as described in Eq. (21), are obtained by using the dense electron-phonon matrix elements obtained from Wannier interpolations.

### C. Electron-impurity scattering

For electrons elastically scattered by ionized impurities, they retain the memory of the incident momentum. Therefore, the scattering is not isotropic and the momentum relaxation time should be considered instead of the average time between scatterings [20]. The inverse of electron relaxation time limited by impurity scattering is formulated as [52]

$$\frac{1}{\tau_{imp}} = \sum_{\mathbf{k'}} S_{\mathbf{kk'}}(1-\cos\theta) . \qquad (31)$$

Here $S_{\mathbf{kk'}}$ is the transition rate between initial and final state $\mathbf{k}$ and $\mathbf{k'}$, respectively. $\theta$ is the angle between the in-coming and out-going momentum. For simplicity, the electron band index is omitted here and for the discussions below. In the framework of Born approximation and by assuming the impurities are uncoupled independent scattering centers, the transition rate can be expressed according to Fermi's golden rule as:

$$S_{\mathbf{kk'}} = \frac{2\pi}{\hbar} N_I V |H_{\mathbf{kk'}}|^2 \delta(\varepsilon_{\mathbf{k}} - \varepsilon_{\mathbf{k'}}) . \qquad (32)$$

Here $V$ is the normalization volume, or the "crystal volume". We have $V = N_{\mathbf{k}} V_{prim}$ where $N_{\mathbf{k}}$ is the number of k-points in the first Brillouin zone resulting from discretization and $V_{prim}$ is the primitive cell volume. $N_I$ is the impurity density with units of cm$^{-3}$. $H_{\mathbf{kk'}}$ is the matrix element describing the strength of electron scattering by a single impurity

$$H_{\mathbf{kk'}} = \langle \mathbf{k'}|\hat{H}|\mathbf{k}\rangle = \frac{1}{V}\int_{\infty} \psi^{(u)*}_{\mathbf{k'}} \Delta U \psi^{(u)}_{\mathbf{k}} d^3r . \qquad (33)$$

Here $\psi^{(u)}_{\mathbf{k}}$ is the wavefunction which satisfies the normalization requirement

$$\frac{1}{V_{prim}} \int_{V_{prim}} \psi^{(u)*}_{\mathbf{k}} \psi^{(u)}_{\mathbf{k}} d^3r = 1. \qquad (34)$$

$\tilde{\psi}_{\mathbf{k}}^{(u)} = e^{i\mathbf{k}\cdot\mathbf{R}}\psi_{\mathbf{k}}^{(u)}$ denotes the Bloch extension of the wavefunction to outside the primitive cell volume $V_{prim}$ and $\psi_{\mathbf{k}}^{(u)}(\mathbf{r}) = e^{i\mathbf{k}\cdot\mathbf{r}}u(\mathbf{r})$, where $u(\mathbf{r})$ is the periodic function of the lattice. It should be noted that QUANTUM ESPRESSO only computes the periodic $u(\mathbf{r})$ part of the eigenfunctions. By substituting Eq. (33) into Eq. (32) and into Eq. (31), the electron-impurity scattering rate can be obtained as

$$\frac{1}{\tau_{imp}} = \sum_{\mathbf{k'}} \frac{1}{N_{\mathbf{k}}} \frac{2\pi}{\hbar} \frac{N_I}{V_{prim}} |\int_{\infty} \tilde{\psi}_{\mathbf{k'}}^{(u)*} \Delta U \tilde{\psi}_{\mathbf{k}}^{(u)} d^3r|^2 \, \delta(\varepsilon_{\mathbf{k}} - \varepsilon_{\mathbf{k'}})(1-\cos\theta). \tag{35}$$

The rate is computed by summing over all possible final states that can satisfy the conservation of energy due to elastic scattering. The $\cos\theta$ is computed by taking the dot product between the group velocities of initial and final states. Here the energy conservation, as represented by the Dirac delta function, is satisfied by searching through the electron band structure computed from DFT. The only quantity left to be evaluated is the impurity scattering matrix element $H_{\mathbf{kk'}}$.

To evaluate $H_{\mathbf{kk'}}$, we first compute the ground-state electron wavefunctions $\psi_{\mathbf{k}}$ from DFT. Then the ionized impurity potential $\Delta U$ needs to be modeled. To model $\Delta U$ due to defects or guest atoms, supercells containing host material and the defect/guest atom are usually constructed. Geometry optimization and ground state self-consistent DFT calculations are done independently to obtain the self-consistent KS potential (including all terms) for both the supercell containing only host material $U_{pure}$ and a supercell containing host material with embedded defects $U_{def}$. The scattering potential due to defects/impurities is taken as the difference between the self-consistent KS potentials of the two supercells $\Delta U = U_{def} - U_{pure}$. This approach works well for charge-neutral defects and other cases where the perturbation from imperfection comes from local lattice strains or charge redistribution.

The supercell treatment is problematic for ionized impurities or defects for which the Coulomb interaction extends to very long range. There are two main issues with this approach: 1) only short range interactions can be treated due to tremendous computational demand. Due to the computational demand of DFT calculations, only a few hundred atoms can be handled, which corresponds to cubic supercells with side length around a few nanometers at most. This corresponds to an effective impurity density of $10^{20}$ cm$^{-3}$ if only one impurity is contained inside the supercell. However, the Debye screening length corresponding to lower impurity densities is clearly beyond the range that can be achieved by constructing supercells. 2) DFT calculation is at ground state. Therefore, the computed KS potential only corresponds to that of the non-ionized impurity. It turns out for substitutional doping of P$^+$ in a Si supercell, the ground-state KS potential due to P$^+$ impurity is highly oscillatory spatially and highly localized [21]. To compute the potential for an ionized impurity, an excited state calculation is likely needed, which is beyond the scope of this work.

Because of afore-mentioned issues, at present, there is no clear way to account for the long range Coulomb interactions due to ionized impurities using only first-principles. In the past, most works on ionized impurity scattering were done empirically with lots of efforts devoted to improving the description of dielectric screening, Born approximation, non-parabollic band, and so on [22]. Recently, Restrepo *et al* [9] and Lordi *et al* [23] modeled the ionized impurities from first-principles but only within a finite supercell. Instead, we choose to use a relaxed approach by using the DFT wavefunctions in combination with the classical model of a screened Coulomb potentials.

The screened Coulomb potential is modeled as:

$$\Delta U = \frac{e^2}{4\pi\varepsilon\varepsilon_0} \frac{1}{r} e^{-r/L_D} \quad . \tag{36}$$

Here $\varepsilon$ is the relative permittivity or dielectric constant, and

$$L_D = \sqrt{\frac{\varepsilon\varepsilon_0 k_B T}{e^2 N_I}} \tag{37}$$

is the Debye length representing the screen length in the case of non-degenerate semiconductors. Here we assumed that doping is in the extrinsic regime and all impurities have been completely ionized. Under such assumption, the carrier concentration equals the impurity density $n_i = N_I$. To evaluate the screened Coulomb potentials, the only unknown is $\varepsilon$, which may be evaluated from first-principles [24]. Therefore the modeling of impurity scattering can still be regarded as *ab initio*. In this work we use commonly accepted values of $\varepsilon = 11.68$ at room temperature for silicon. The effect of lattice strain due to different dopants was found to affect the mobility in a systematic way [25]. The effect of choice of different dopants is omitted in this work for simplicity.

With the impurity potential modeled, yet another major challenge is to evaluate the real space integral required to evaluate the matrix element, as in Eq. (33). Due to the oscillatory feature of the wavefunction and the long range tail of the impurity potential $\Delta U$, a fine grid extending beyond at least ten times of the Debye length $L_D$ is typically required for converged integral. Such integration turns out to be numerically difficult. Instead, we note that for Bloch waves, the integral in Eq. (33) can be approximately decomposed following the approach outlined in Ref. [20,26]:

$$H_{\mathbf{kk'}} = I(\mathbf{k},\mathbf{k'})U_s(\mathbf{k}-\mathbf{k'}) \quad . \tag{38}$$

Here $I(\mathbf{k},\mathbf{k'}) = \int_{V_{prim}} u_{\mathbf{k'}}^*(\mathbf{r})u_{\mathbf{k}}(\mathbf{r})\,d^3r$ is the overlap integral that can be readily computed by integrating the DFT wavefunctions over the real space volume of a single primitive cell. On the other hand,

$$U_s(\mathbf{k}-\mathbf{k}') = \int_\infty \frac{e^{-i\mathbf{k}'\cdot\mathbf{r}}}{\sqrt{\Omega}} \Delta U \frac{e^{i\mathbf{k}\cdot\mathbf{r}}}{\sqrt{\Omega}} d^3r \ , \tag{39}$$

is nothing but the planewave solution of the electron-impurity matrix elements. It has analytical solution as

$$U_s(\mathbf{k}-\mathbf{k}') = \frac{e^2}{V\varepsilon\varepsilon_0} \frac{1}{|\mathbf{k}-\mathbf{k}'|^2 + 1/L_D^2} . \tag{40}$$

Therefore the electron-impurity matrix element can be simplified as

$$H_{\mathbf{kk}'} = \frac{e^2}{V\varepsilon\varepsilon_0} \frac{I(\mathbf{k},\mathbf{k}')}{|\mathbf{k}-\mathbf{k}'|^2 + 1/L_D^2} , \tag{41}$$

which is the DFT overlap function multiplied by the planewave solution of impurity scattering.

With the matrix element known, Eq. (35) is used to compute the electron-impurity scattering rate. The dependence of impurity density enters both through the multiples of single impurity scattering rate and through the Coulomb screening effect in Debye length in the matrix element $H_{\mathbf{kk}'}$.

### D. Phonon-phonon scattering

The phonon-phonon scattering rate is computed based on the anharmonic lattice dynamics. More detailed introduction can be found elsewhere [4]. In short, the harmonic and cubic force constants are obtained by fitting to the force-displacement relations computed using DFT for a selection of supercells [27]. Based on the fitted harmonic force constants, phonon dispersions can be obtained by solving the dynamical matrix. The inverse of phonon relaxation time relates to the imaginary part of the three-phonon self-energy as [4]: (are you missing h in expression below?)

$$\frac{1}{\tau_{\mathbf{q}\nu}^{ph}} = 2\,\mathrm{Im}[\Sigma(\mathbf{q}\nu, \omega_{\mathbf{q}\nu})] \ . \tag{42}$$

By applying Fermi's golden rule to the cubic Hamiltonian, the imaginary part of the three-phonon self-energy can be obtained as [4]:

$$\Sigma(\mathbf{q}\nu,\omega) = -\frac{1}{2\hbar^2} \sum_{\mathbf{q}_1\nu_1, \mathbf{q}_2\nu_2, \varepsilon=\pm 1} w_{\mathbf{q}} |U(\mathbf{q}\nu, \mathbf{q}_1\nu_1, \mathbf{q}_2\nu_2)|^2$$
$$\times \left[ \frac{1+n_1+n_2}{\omega_1+\omega_2+\varepsilon(\omega-i\eta)} + \frac{n_2-n_1}{\omega_1-\omega_2+\varepsilon(\omega-i\eta)} \right] . \tag{43}$$

Here $w_{\mathbf{q}} = 1/N_q$ is the weighting factor and $\eta$ is again the Gaussian broadening parameter. The phonon-phonon matrix element is

$$U(\mathbf{q}\nu,\mathbf{q}_1\nu_1,\mathbf{q}_2\nu_2) = \left(\frac{\hbar}{2}\right)^{3/2} \sum_{R_i b_i \alpha_i} \Psi^{\alpha\beta\gamma}_{0b_0, R_1 b_1, R_2 b_2}$$
$$\times \frac{2^{i(\mathbf{q}_1 \cdot \mathbf{R}_1 + \mathbf{q}_2 \cdot \mathbf{R}_2)} e_{b_0 \alpha}(\mathbf{q}\nu) e_{b_1 \beta}(\mathbf{q}_1 \nu_1) e_{b_2 \gamma}(\mathbf{q}_2 \nu_2)}{\left(m_{b_0} m_{b_1} m_{b_2} \omega_{q\nu} \omega_1 \omega_2\right)} \quad (44)$$

Conservation of momentum requires $\mathbf{q}_1 + \mathbf{q}_2 = \mathbf{G} - \mathbf{q}$. $\Psi^{\alpha\beta\gamma}_{0b_0, R_1 b_1, R_2 b_2}$ is the cubic force constant and $e_{b_i \alpha_i}(\mathbf{q}\nu)$ is the $\alpha_i$ th component of the part corresponding to the $b_i$ th basis atom in the eigenvector of phonon mode $\mathbf{q}\nu$.

With the knowledge of cubic force constants and harmonic phonon properties, the phonon-phonon matrix element can be evaluated. The phonon self-energy, and thus the phonon relaxation time can be obtained by searching through all possible phonon-phonon scattering events satisfying both momentum and energy conservation.

## II. Numerical results for scattering rates

### A. DFT calculations

The QUANTUM ESPRESSO package [15] is used to perform all DFT calculations for electrons and DFPT calculation for phonons. The local density approximation (LDA) is used for the exchange and correlation functional. The KS equations are solved using the standard pseudopotential and plane-wave approaches as implemented in QUANTUM ESPRESSO. For silicon, we use the norm-conserving LDA potentials with Perdew-Zunger data. The plane-wave basis cutoff was set to be 48 Ry. Shifted uniform Monkhorst-Pack grids with size of $16 \times 16 \times 16$ are used for all self-consistent calculations (scf) and lattice relaxations to achieve full convergence. The fully relaxed lattice constant is found to be 5.4 Å.

From the computed electronic bandstructure, the predicted indirect electronic bandgap in silicon is found to be 0.43 eV at 0 K, which is significantly underestimated from that observed experimentally. Such underestimation is common due to the use of LDA approximation and general gradient approximation (GGA) that tends to treat the system as metallic [28]. Other more sophisticated approaches such as GW method can be used to predict the correct bandgap in many materials. Here, rigid band approximation will be assumed to shift the entire valence band uniformly to open up the bandgap to match experimentally observed values about 1.12 eV. As can be seen later, this rigid band approach is reasonable enough for the transport calculation of electrons.

### B. Electron-phonon scattering
#### a. Chemical potential inside bandgap

To compute the electron-phonon scattering rate, we use the EPW package [18], which is able to perform Wannier interpolation using the DFT outputs from QUANTUM ESPRESSO. To better suit our needs, we have modified the EPW package. By following the methodology outlined in the previous sections, we obtain the electron-phonon scattering rate for electronic states near the conduction band minimum (CBM). For simplicity, only electrons are considered in this work although holes can be treated in a similar fashion.

The electronic states, phonon dispersions and electron-phonon matrix elements are first calculated on coarse electron and phonon uniform grids both at $8\times8\times8$. Then MLWFs are constructed and used to interpolate all quantities to a dense uniform grid of $30\times30\times30$ for electrons and $60\times60\times60$ for phonons, respectively. Such dense grids are found to lead to converged electron-phonon scattering rate with degauss parameter choice of 0.03 eV to account for the energy conservation requirement enforced by the Dirac delta functions. A typical parallel run of electron-phonon scattering calculation using 16 cores takes about 12 hours on a computer cluster with Intel Xeon 2.6 GHz CPUs. We will not categorize the scattering processes into intervalley and intravalley scatterings since they have already been abundantly discussed in literature for silicon [29].

When the chemical potential is chosen to be deep inside the bandgap, the resulting electron-phonon scattering rate corresponds to that of intrinsic silicon. The computed total electron-phonon scattering rate and its decomposition into contributions from different phonon branches are shown in Fig. S1. The spread of data around a given energy is due to the anisotropy of scattering in the first Brillouin zone. As seen in Fig. S1 a), starting from the CBM, the scattering rate increases with electron energy and follows the general trend of the electron density of states (DOS). In general, for electrons at higher energy states, the phase space of possible final states satisfying energy and momentum conservation for scattering is much larger. Therefore, scattering becomes more frequent, leading to higher scattering rates. Our computed scattering rate based on Wannier interpolation shows reasonable agreement with previous DFT [9,17] and tight-binding calculations [30] with linearly interpolated electron-phonon matrix elements. The deviation in values, especially at low energies, from previous DFT calculations is attributed to the use of very dense grids in this work. Nonetheless, the apparent overall agreement is expected for silicon, in which the energy dependence of the matrix elements can be reasonably described by simple linear dispersion. However, this may not be true for more complicated thermoelectric materials.

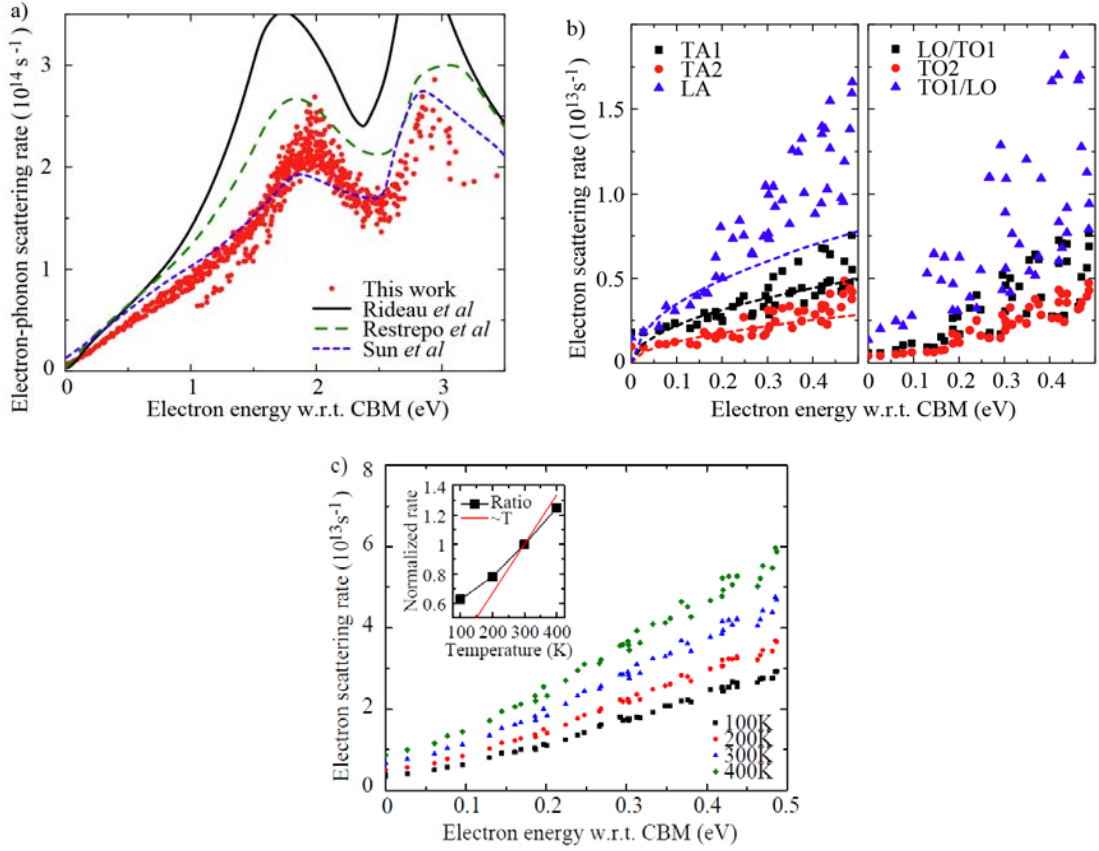

Figure S1. a) Electron-phonon scattering rate at 300 K in comparison with results from previous DFT with linear interpolation (dashed curves) [9,17] and tight-binding (solid curve) [30]. b) Scattering rate decomposition into different phonon branch contributions. Dashed-lines are trend curves indicating expected energy dependence of $\tau^{-1} \propto E^{1/2}$. c) The temperature dependence of electron-phonon scattering rate. Inset shows the average value of normalized scattering rate w.r.t. the rate at 300 K and the trend curve indicates $\tau^{-1} \propto T$.

The acoustic and optical phonon contribution to electron-phonon scattering is shown in Fig. S1 b). We can first see that, for the energy range most relevant to the electron transport, longitudinal acoustic (LA) phonons scatter strongly with electrons while the two transverse acoustic (TA) phonons scatter electrons with similarly lower strength. The energy dependence of the scattering rates, however, deviates from the $\tau^{-1} \propto E^{1/2}$ dependence expected from deformation potential theories at high energies [20]. Particularly, the deviation is strongest above ~ 0.2 eV. In fact, if we replace the energy dependent electron-phonon matrix elements with averaged constant values in the same spirit of deformation potentials [14], the resulting energy dependence of the scattering rate recovers the power law. Therefore, the deviation in our results from earlier theories indicates the necessity of taking into account the complicated dispersions of electron-phonon matrix elements and non-parabollic band structures instead of assuming averaged constant deformation potential and parabolic bands. This also explains the wide spread of deformation potential values reported in literature. For optical phonon contributions, because of the band-crossing between longitudinal optical (LO) and first transverse optical (TO1) phonon branches throughout the Brillouin zone, we report the resulting scattering rates in a mixed manner instead. As compared to acoustic phonon scattering, the scattering by optical phonons is relatively weak at lower electron energies but becomes comparable in strength as electron energy increases.

The temperature dependence of the scattering rate is shown in Fig. S1 c). As temperature increases, phonon population increases so that electrons experience more frequent scattering by phonons, leading to higher scattering rate. Since the phonon population is roughly proportional to temperature T at high temperatures, it might be expected that the scattering of electrons should follow the same temperature dependence. However, as shown in the inset of Fig. S1 c), when we average the scattering rates and normalize it using the value at 300 K, we found the temperature dependence deviate from the linear $T$ dependence. Especially, at lower temperatures the scattering is higher than expected. In fact, when the chemical potential is inside the bandgap, the electronic states in the conduction band are less occupied at lower temperatures. Therefore, there are more empty states for electrons to scatter into. As a result, scattering rate is higher compared to when we assume the electron population is irrelevant.

b. Chemical potential inside conduction band

The electron-phonon scattering rates as a function of temperature when the chemical potential is at 0.27 eV above the CBM is shown in Fig. S2. At room temperature, such chemical potential corresponds to the carrier concentration of $6\times10^{20}$ cm$^{-3}$, which is in a range that is feasible in highly doped silicon [31]. To achieve enough resolution, scattering rates are recomputed on a uniform electron grid of $60\times60\times60$. Around the chemical potential $\mu$, there is clear dip in the scattering rates. Such a dip is strong at low temperatures. Considering the narrow Fermi window, only electrons with energies about less than $3k_BT$ away from the chemical potential contribute significantly to the transport, the energy dependence of the scattering rate near the chemical potential will dominate the transport. With the chemical potential deep inside the conduction band, the electron carrier concentration is so high that silicon approaches the metallic regime. For simple metals, by assuming a parabolic bandstructure of $E \propto k^2$ that is characteristic for free electrons, the rate of electrons scattered by acoustic phonons is predicted to exhibit the energy dependence of $\tau^{-1} \propto E^{-3/2}$ [1]. However, as seen in Fig. S2 a), there is no resemblance of the shapes of energy dependence around the chemical potential to that of the power law dependence. This disagreement is likely due to the fact that the electron bandstructure around the chemical potential is not parabolic. As a result, the power law energy dependence that relies on the assumption of parabolic bandstructure does not apply. In fact, the free electron parabolic energy dispersion is only true for simple metals such as alkalis.

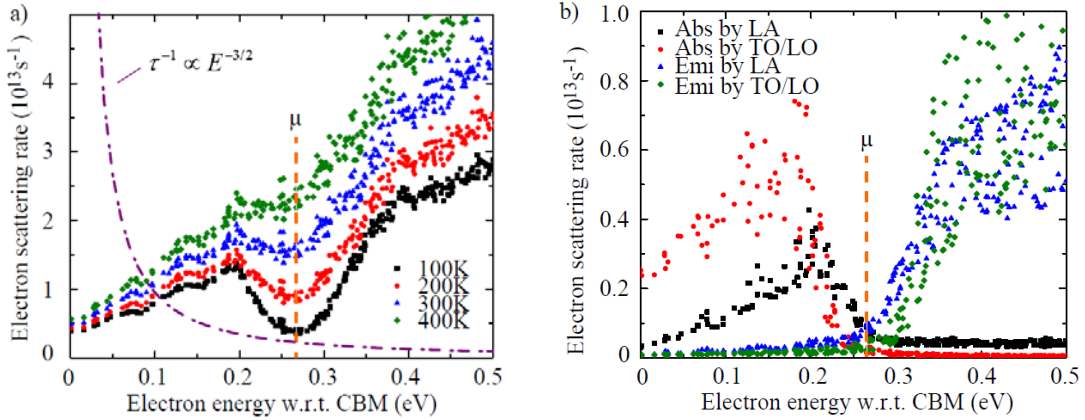

Figure S2. a) Electron phonon scattering rate when the chemical potential $\mu$ is positioned inside the conduction band of silicon. The dash-dotted curve represents metal theory where $\tau^{-1} \propto E^{-3/2}$. b) The absorption (Abs) and emission (Emi) rates due to LA and LO/TO phonons at 100 K.

In order to see the origin of the dip, we decompose the scattering rate into absorption and emission processes. For clarity, we only show scattering rates due to LA and TO/LO phonons in Fig. S2 b), while the trends are very similar for contribution from other phonon branches. As seen, the absorption rate rapidly declines as electron energy approaches the chemical potential from the lower end. In contrast, the emission rate rapidly declines as electron energy approaches the chemical potential from the higher end. The decline is faster for optical phonons than acoustic phonons. Such change in scattering rate is due to the selection rule and the occupancy of electron states. If we look at the emission rate, the scattering process involves an initial electronic state emitting a phonon to jump to a final electronic state with lower energy. For states with energy much higher than the Fermi level, their adjacent states are mostly empty and the scattering can easily happen. However, for states just above the chemical potential, they cannot emit a phonon and jump to a lower energy state because all possible final states are essentially occupied. Such forbidden scattering is particularly evident at lower temperatures since the Fermi-Dirac distribution resembles a step function. Similar arguments also apply to the absorption process. As a result, the overall scattering rate as a sum of absorption and emission processes shows a dip around the chemical potential. Because of the smearing of Fermi-Dirac distribution at higher temperatures, such dip in scattering rate is only important at low temperatures. Similar dip in scattering rate has also been observed in graphene at low temperatures [32].

### C. Electron-impurity scattering

The task of computing the electron-impurity matrix elements is reduced to computing the overlap integral of crystal wavefunctions. To ensure the appropriate resolution within the small energy range above the CBM that is relevant to transport calculations, we constructed a dense uniform electron grid of $100 \times 100 \times 100$. To reduce the computational cost, only those electronic states with energies within 0.1 eV above the CBM are computed for scattering rate while the search for possible scattering events extends to 0.5 eV above the CBM. The degauss parameter used in Gaussian broadening treatment of Dirac delta function is found to converge at 0.0015 eV. We also constructed a coarse uniform grid of $36\times36\times36$ for computing electronic states with energy up to 0.5 eV above the CBM to confirm the trend and convergence with respect to the grid density. We assume fully ionized impurities so the carrier concentration equals that of the impurity density. These two terms will be interchangeable throughout the rest of this article.

The electron-impurity scattering rates are computed on both dense and coarse grids following the methodology outlined in the previous sections. For clarity, only the results from impurity density $n_i = 10^{18}$ cm$^{-3}$ are shown in Fig. S3 a). Similar to the case of electron-phonon scattering, the raw impurity scattering rate data spreads for given electron energies. This is again due to the anisotropic scattering in the first Brillouin zone. Despite the spread of data, it is clear that the results from both dense and coarse grid fall into the same trend, indicating the convergence of the rate with respect to grid density. In general, the scattering rate is high for low electronic energy states and it monotonically decreases as electron energy increases. This can be understood for electrons with low energies since they can probe the perturbative potential created by the ionized impurities; consequently, the probability of scattering is much higher. For electrons with high enough energy, the perturbative potential is negligible, resulting in a low scattering rate.

As seen in Fig. S3 a), the energy dependence of the electron-impurity scattering rate can be generally described by a power law dependence $\tau^{-1} \propto E^{-\alpha}$. By fitting to the raw scattering data for impurity densities ranging from $10^{16}$ to $10^{19}$ cm$^{-3}$, we have obtained the exponents $\alpha$ as listed in Fig. S3 b) along with the fitted curves. The exponent spans from 3.6 to 0.5, depending on impurity density. It was predicted from the previous theory that for free electron scattering from a weakly screened ionized impurity, the energy dependence of the scattering rate should follow $\tau^{-1} \propto E^{-3/2}$ power law [20]. The deviation in the exponent is due to the dielectric screening of the charged impurities and the overlap integral between crystal wavefunctions inherent to bound electrons. For simplicity in the later computation of transport coefficients, instead of directly using the raw impurity scattering rate data, we will only use the best fit curves shown in Fig. S3 b) to represent impurity scattering.

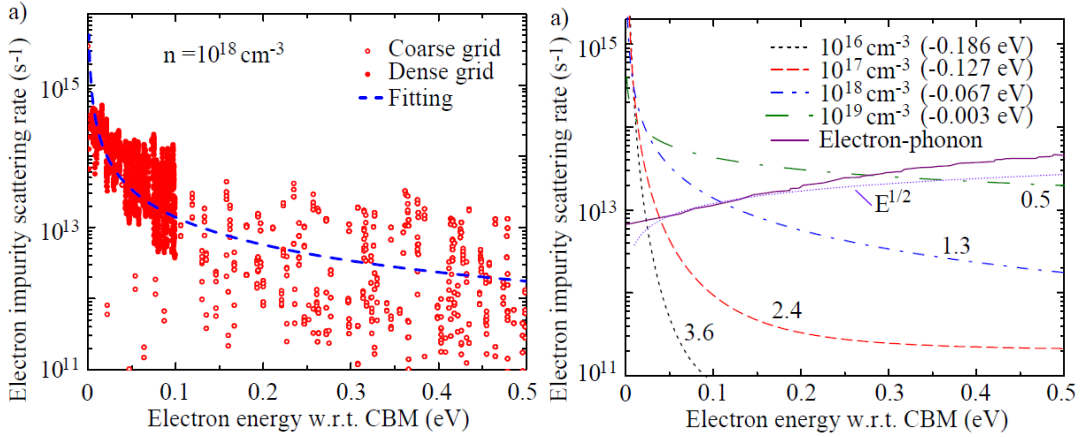

Figure S3. Electron-impurity scattering rates at 300 K. a) The raw electron-impurity rate at impurity density of $10^{18}$ cm$^{-3}$ from both the coarse and dense grid. The dashed curve is the best fit. b) The fitted electron-impurity rate for a range of impurity densities in comparison to that of electron-phonon scattering rate. The numbers near the curves indicate the exponent $\alpha$ in $\tau^{-1} \propto E^{-\alpha}$.

The scattering rate is generally found to increase with the impurity density, as shown in Fig. S3 b). However, we find the predicted scattering rate saturates and even begins to decrease when the impurity density exceeds $10^{20}$ cm$^{-3}$ (not shown). This observation is somewhat unexpected but can be interpreted as the competition between the reduced Debye screening length and the increasing impurity density. Since we have made the assumption to treat impurities as independent scattering centers, the scattering rate always has a prefactor that is directly proportional to the impurity density. On the other hand, as impurity density increases, due to stronger dielectric screening, the Debye length decreases accordingly. Such shorter Debye length indicates that the scattering strength from individual scattering center is decreasing with increasing impurity density. Therefore, at high enough impurity density, the impurity scattering rate appears to decrease. This is however unphysical and due to the breakdown of the assumption of independent scattering centers. In fact, at very high impurity densities, the coupling among ionized impurities becomes important and cannot be neglected. The explicit treatment of coupling among impurities can be tedious and is under investigation. In addition, electron-electron scattering may become important at very high carrier concentration. This is neglected in the present work for the range of impurity densities under consideration.

As seen in Fig. S3 b), in comparison to electron-phonon scattering rates, impurity scattering rates dominate the low energy regime. As impurity density increases, larger energy span from the CBM is being governed by impurity scattering. Electron-phonon scattering is only important at high electron energies or high temperatures, depending on the impurity density. This is consistent with the finding that at high field transport in electronic devices, it is the electron-phonon process that directly leads to heat generation dominates the transport [33].

## III. Transport coefficients

With the total electron scattering rate computed, the electron relaxation time is obtained as its inverse. The energy dependent relaxation times and the electronic bandstructure computed from DFT are then used to obtain the electron transport coefficients, following the methodology outlined in previous sections. All coefficients are computed using the BOLTZTRAP package [8], in which we modified the code to incorporate energy dependent relaxation time models and corresponding interpolation instead of using the default CRTA. We further assume the introduction of ionized impurities will only introduce additional carriers, which shifts the chemical potential without modifying the bandstructure itself. Such approximation is reasonable for low to intermediate doping while it may become invalid at significant doping concentrations. In fact, it was suggested that at very high doping concentrations, the impurities will couple strongly and interact with host lattice to form delocalized impurity bands. These strong interactions introduce so called resonant states and hence modify the electronic bandstructure and transport coefficients [34].

With rigid band approximation, the chemical potential is related to the carrier concentration as:

$$n_{net} = n_{d,h} - n_{d,e} = \int_{-\infty}^{-E_g} D(\varepsilon) \frac{1}{e^{(\mu-\varepsilon)/k_BT}+1} d\varepsilon - \int_{0}^{\infty} D(\varepsilon) \frac{1}{e^{(\varepsilon-\mu)/k_BT}+1} d\varepsilon. \tag{45}$$

Here $D(\varepsilon)$ is the electron DOS, $\mu$ is the chemical potential, and $E_g$ is the bandgap. For clarity, we only look at the n-type silicon where the chemical potential is close to the CBM. Then, in the extrinsic regime, the transport coefficients are not sensitive to the choice of bandgap in silicon. Nevertheless, we correct the bandgap with experimentally observed value of 1.12 eV by rigidly and uniformly shifting the valence bands to lower energy. For a given carrier concentration and temperature, the position of the chemical potential can be determined by solving Eq. (45). Then the transport coefficients can be obtained at the prescribed chemical potential from Eq. (14).

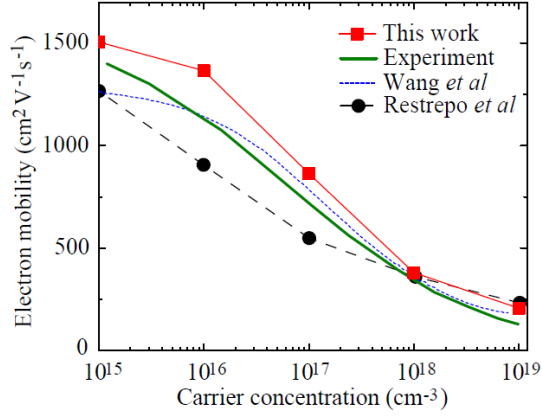

Figure S4. Electron mobility in silicon at 300 K. Electron mobility from experiment [37] and previous DFT works [7,9] are shown for comparison.

The computed electron mobility at 300 K is shown in Fig. S4. For carrier concentrations below $10^{15}$ cm$^{-3}$, the mobility is mainly limited by electron-phonon scattering. For higher carrier concentrations, the mobility decreases due to increased strength in electron-impurity scattering. Though not shown, the mobility will decrease with increasing temperature mainly due to stronger electron-phonon scattering. Additionally, our predicted carrier concentration dependence of the electron mobility agrees with experiments reasonably well. Also shown in Fig. S4 are previous DFT-based mobility calculations. Specifically, Wang *et al* computed deformation potentials from DFT for electron-phonon scattering then used empirical models for impurity scattering [7]. Restrepo *et al* used linear interpolation for electron-phonon matrix elements and modeled ionized potential within a supercell [9]. We speculate the difference among our work and previous DFT works to be due to the use of Wannier interpolated dense grid and the inclusion of long range Coulomb tails due to ionized impurity, and the use of DFT wavefunction in the computation of matrix elements.

The Seebeck coefficients as a function of carrier concentration at elevated temperatures are computed from Eq. (14), as shown in Fig. S5. The Seebeck coefficients decrease with increasing carrier concentration, which is consistent with the fact that Seebeck is highest in insulators while lowest in metals. In the inset of Fig. S5, it is seen that for the carrier concentration under consideration, the Seebeck coefficient increases with increasing temperature. For a given carrier concentration, as temperature increases, the chemical potential shifts away from CBM due to the smearing of Fermi-Dirac distributions. Therefore the average energy of carriers characterized by $\varepsilon - \mu$, as in Eq. (14), increases, leading to increased Seebeck coefficient.

For non-degenerate semiconductors, one can approximate the Fermi-Dirac distribution of electrons as the Boltzmann distribution. Then also by assuming a parabolic bandstructure for electron DOS, the Seebeck coefficients for electrons can be approximated by solving Eq. (14) as [1]:

$$S = -\frac{k_B}{e}\left(\frac{5}{2} + \alpha - \ln\frac{n}{N_c}\right). \tag{46}$$

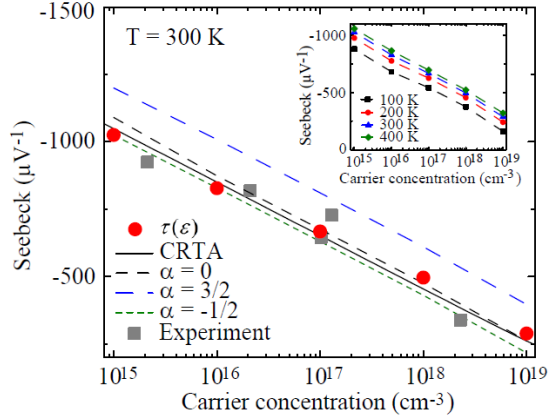

Figure S5. Seebeck coefficients as a function of carrier concentration at 300 K in comparison to those from approximations with different choices of exponent $\alpha$. Experimental data from Ref. [7] which has excluded the effects of phonon drag from Ref. [38]. The inset shows the temperature dependence of Seebeck coefficients.

Here $\alpha$ is the exponent originating from the electron scattering rate $\tau^{-1} \propto E^{-\alpha}$. The inclusion of this exponent refers to the dependence of Seebeck coefficients on the particular scattering mechanisms. $n$ is the carrier concentration and $N_c$ is the effective density of states. The estimation using Eq. (46) with different choices of exponent $\alpha$ is shown in comparison to the DFT results at 300 K. As seen, the DFT results fall between $\alpha = -1/2$ and $\alpha = 0$. This is a strong deviation from the $\alpha = 3/2$ dependence suggested by previous theories of impurity scattering. This can be understood from the combined electron scattering rate, as shown in Fig. 4. Throughout the energy range relevant for electron transport, the energy dependence of the scattering rate exhibits a transition from impurity scattering to phonon scattering, depending on impurity densities. Therefore, the effective exponent $\alpha$ appears to be a weighted average over the energy span, leading to a value not very different from 0. It should be noted that when $\alpha = 0$, the energy dependence of scattering rate becomes $\tau^{-1} \propto E^0$, which is basically the CRTA approximation. In Fig. S5 we also show the Seebeck coefficient directly computed from BOLTZTRAP with default CRTA, which agree reasonably well with the DFT results with fully energy dependent relaxation time $\tau(\varepsilon)$. Therefore, due to the transition of dominating electron scattering mechanisms within the Fermi window, the CRTA is an effective approximation for simple materials such as silicon for predicting Seebeck coefficients.

The electrical conductivity and electron thermal conductivity as a function of carrier concentration at elevated temperatures are computed from Eq. (14), as shown in Fig. S6 a). As expected, both electrical conductivity and electron thermal conductivity increase with increasing carrier concentrations. Also, at temperatures above 100 K, the electrical conductivity decreases with increasing temperature for a given carrier concentration. This is mainly due to the decrease in electron mobility owing to stronger electron-phonon scattering that is most evident at lower carrier concentrations. The deviation from the linear dependence of electrical conductivity on carrier concentration at concentrations larger than $10^{16}$ cm$^{-3}$ is also due to the decrease in electron mobility as carrier concentration increases.

The electronic thermal conductivity remains small even at carrier concentrations as high as $10^{19}$ cm$^{-3}$. For metals, the Wiedemann-Franz law well describes the relation between electrical conductivity and electronic thermal conductivity as:

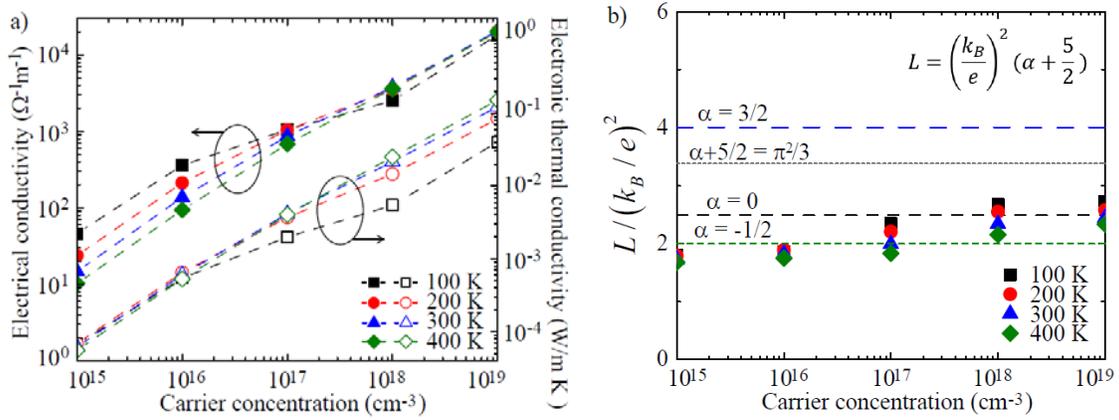

Figure S6. a) Electrical conductivity and electron thermal conductivity as function of carrier concentration and temperature. b) The normalized Lorenz number as a function of carrier concentration.

$$k^{el} = L_0 \sigma T. \qquad (47)$$

Here $L_0 = \dfrac{\pi^2}{3}\left(\dfrac{k_B}{e}\right)^2 = 2.45\times 10^{-8}\,(\text{W }\Omega\text{ K}^{-2})$ is the Lorenz number. For semiconductors, there may be deviations in the Lorenz number from that of metals. In fact, for non-degenerate semiconductors we can also assume parabolic bandstructure and power law dependence of electron scattering rate. We can then solve Eq. (14) and express the Lorenz number as [35]:

$$L = \left(\dfrac{k_B}{e}\right)^2\left(\alpha + \dfrac{5}{2}\right). \qquad (48)$$

To see the applicability of Wiedemann-Franz law, we compute the Lorenz number using our DFT results by directly taking the ratio between $k^{el}$ and $\sigma$, as shown in Fig. S6 b). Similar to what we have observed for Seebeck coefficients, the computed Lorenz number falls into the range of $\alpha = -1/2$ to $\alpha = 0$. This is again due to the mixed energy dependence of the electron scattering rate. Also, the obtained Lorenz number is found to be smaller than in metal by a ratio of $L/L_0 \approx 0.76$. This is because silicon is still non-degenerate for the carrier concentration range under consideration. Depending on the electron scattering rate and electronic bandstructure in different non-degenerate materials, $L$ can also be larger than $L_0$ [36].